# Cardiopulmonary responses and muscle strength influence running performance parameters differently at different slopes

Corentin Hingrand[1]*, Adrien Combes[2], Nicolas Olivier[2], Samir Bensaid[2], Frédéric N. Daussin[2]

[1] Normandie Univ, UNICAEN, VERTEX UR 7480, 14000 Caen, France
[2] Université de Lille, Université d'Artois, Université de Littoral Côte d'Opale, ULR 7369 - URePSSS-Unité de Recherche Pluridisciplinaire Sport Santé Société, Lille, France.

*Corresponding author
Correspondence: corentin.hingrand@unicaen.fr

**ABSTRACT**

The analysis of trail-running performance appears to be complex and cardio-respiratory and muscular factors could have a variable importance depending on the inclination. Our study aims to determine the role of these parameters in performance. 13 subjects with heterogeneous levels participated in the study. They carried out 7 visits including 3 maximal aerobic speed (MAS) test at 1, 10 and 25% slope on treadmill, 3 endurance tests at 100% of the MAS reached at 1, 10 and 25% and an evaluation on isokinetic ergometer at different speeds (60-180-240 °/s). Gas exchange measured during the incremental tests. We were able to identify 2 groups, a performance and a recreational group. We observe a difference in VO2max, MAS at 1 and 10%, and maximal aerobic ascensional speed (MAaS) at 25%, between the 2 groups but no difference in VO2max and exhaustion time at 100% MAS between the different conditions (1-10-25%). Interestingly, at ventilatory thresholds the metabolic parameters, expressed as absolute or relative values, are similar between conditions (10-25%) while the ascensional speed are different. This study suggests that the measurement of ascensional speed is not as relevant as heart rate for controlling intensity given the variety of slope gradients in trail-running races.

## Introduction

Trail running is defined by the International Trail Running Association (ITRA) as a walking competition open to all, in a natural environment, with the minimum possible amount of asphalt roads (20% maximum). The course can be a few kilometres long for the short distances to well over 80km for the ultra-trails, and includes flat, uphill, and downhill sections.

Different physiological parameters are used to characterise trail running athletic performance. Ehrström et al. developed a physiological variable based mathematical model that predicts trail-running performance ($r^2$=0.98) **(McLaughlin et al. 2010; Ehrström et al. 2018)**. In that model, maximal oxygen uptake (VO2max) appears to be less influential on performance (r=0.76) compared to muscle endurance (r=0.91). Moreover, the percentage of maximal aerobic speed (MAS) maintained over a 75km trail run and isometric maximal voluntary knee extensor strength are correlated **(Ehrström et al. 2018)**. With a plurality of running profiles and the multiplication of environmental parameters, effort management – the runner's ability to repeat maximal efforts while reproducing a stable level of performance associated with muscular endurance and VO2max – appears to be the determining factors of performance in competition **(Kerhervé et al. 2015; Balducci et al. 2016; Ehrström et al. 2018)**. Therefore, measurement of VO2max and muscular endurance needs to be performed to characterise a trail running athlete.

The traditional test for assessing aerobic capacity performed on the flat does not seem relevant in prescribing uphill or downhill training sessions **(Ehrström et al. 2018)**. The recommended gradients would then be in the order of 15 to 20% **(Billat 2005)**. Previous studies have shown different VO2max values for untrained endurance subjects, depending on the incline or horizontality of the test (**(Taylor et al. 1955; Astrand and Saltin 1961)**. However, for trained and trail runner subjects, the results are contradictory. Some studies have observed different VO2max **(Taylor et al. 1955; Astrand and Saltin 1961; Paavolainen et al. 2000; Scheer et al. 2018)** between the flat and incline test, while others did not **(Davies et al. 1974; Kasch et al. 1976; Balducci et al. 2016)**. The slope percentage is one parameter that influences VO2max values (**(Paavolainen et al. 2000; Scheer et al. 2018)**. The lowest VO2max values were observed on a traditional treadmill test with 1% inclination, whereas the highest values were observed on an incremental test with an increase of 0.5 km/h and 1% additional inclination every minute. In 2017, Doucende et al. validated the first trail specific test with a 25% incline slope **(Doucende et al. 2022)**. This test measures maximal aerobic ascensional speed (MAaS), and is a good indicator of the ability to run uphill. However, peripheral fatigue could limit the effort of runners not trained on this type of slope, and lowers the maximal running speed achieved. The influence of slope on physiological responses needs to be clarified in trail-runners.

Therefore, the aim of this study is to investigate the influence of slope inclination on performance (maximal aerobic speed and maximal aerobic ascensional speed), cardiorespiratory performance (VO2max, ventilatory threshold, heart rate), and muscular responses.

## Methods

This study was filed with the Research Ethics Committee of the University of Lille under number 2019-381-S77, and received approval on 10 December 2019.

### Participants

In total, 13 active runners (9 men and 4 women, 26 ± 8 years old, 1.72 ± 0.11m, 64 ± 12 kg, 13 ± 6% body fat, 21±2 BMI) were recruited based on the following criteria: practice of trail running in competition at least twice a week and proof of participation in at least one competition in the year prior to inclusion. Any athlete who had an injury that forced them to stop training for at least one week in the previous six months was excluded. All subjects signed an informed consent form. Our study complied with ethical rules and the Declaration of Helsinki.

Procedure

The study took place at the University of Lille, in the Eurasport facility. The study consisted of seven visits. The first visit consisted of an isokinetic assessment of the knee extensors and flexors. The second, fourth, and sixth visits consisted of a MAS test that measured cardiorespiratory parameters on a treadmill at a 1, 10, or 25% slope in a random order. The third, fifth, and seventh visits consisted of a time to exhaustion test at a 1, 10, and 25% slope at 100% of the respective MAS with a heart rate measurement.

Isokinetic evaluation

The knee extensors and flexors were tested during the isokinetic evaluation on CON-TREX Multi-joints (Physiomed, GER). The first test consisted of three successive repetitions at 60°/s to measure the peak force torque in extension and flexion. The second test consisted of six successive repetitions at 180°/s to measure the relative power. The last test consisted of 30 successive repetitions at 240°/s to measure relative power on the last three repetitions and fatigue index. The dominant and non-dominant legs were tested in each test.

MAS determination

MAS tests were performed under three randomized conditions: a 1, 10, and 25% slope on a treadmill (Intelligent Suspension 3 treadmill, Cybex International Inc, MA, USA). The starting increment was 8, 4, and 2 km/h respectively for the 1, 10, and 25% conditions. The fixed increment was 1 km/h at 1%, 1 km/h at 10%, and 0.4km/h at 25%. Oxygen consumption was measured with a gas exchange analyzer (Metamax®, Cortex, Leipzig, Germany) and the subject's heart rate with a heart rate monitor (Polar® H10, Polar Electro Oy, Finland). Ventilatory thresholds were determined graphically as previously described (11).

Time to exhaustion test

After a 15-minute warm-up, runners had to maintain 100% of the MAS reached at a 1, 10, or 25% slope for as long as possible. Heart rate was recorded with a heart rate monitor (Polar® H10, Polar Electro Oy, Finland).

Rating Perceived Effort

At the end of the six running tests, we also evaluated the perception of effort by the rated perceived exertion (RPE) scale of Borg, and the muscular perception by the CR-10 scale of Borg (4).

Statistical Analysis

Data are presented as mean ± standard deviation. The following analyses were performed using Prism software (version 8.0 2018, GraphPad Software2365, San Diego, CA, USA) for each of the three conditions: testing for normality and homogeneity of variance and 2-way ANOVA for each parameter allowed us to compare differences between conditions and groups. Relationships between performance and strength and endurance variables were determined using a Pearson correlation. Significantly associated variables were entered into a linear regression model as independent variables. Before determining the multiple linear regression, the multi-collinearity of the predictors was checked. The significant threshold was defined as $p < 0.05$.

**Results**

Results are presented either using all subjects or using two groups of level determined using the ITRA index. For the men, the barrier was set at a score higher than 600, and for the women a score higher than 550. This enabled us to split the group into two sub-groups: a performance group (ITRA index: 679.7 ± 66.14) and a recreational group (ITRA index: 446.2 ± 90.6).

There was no difference between the dominant and non-dominant leg for each group either for the peak force torque in leg extension and flexion at 60°/s, or for the relative power at 180°/s and 240°/s in leg extension and flexion. Similarly, there was no difference between the trained and less trained groups for the different parameters measured, expressed in absolute or relative values.

MAS decreased for all subjects with the slope (18 ± 2.5km/h vs 13 ± 1.7km/h vs 6.7 ± 1 km/h; at the 1, 10, and 25% slope, respectively, $p<0.001$, Fig. 1.B). MAaS increased for all subjects with the slope ($p<0.001$), and were different between the performance group vs recreational group (10% slope: 1371.4 ± 75.6 vs 1133.4 ± 163.3; 25% slope: 1785.7 ± 203 and 1533.4 ± 250 m/h, respectively, $p<0.05$, Fig 1. C). No effect of the slope on VO2max was assessed during the incremental test, as was observed for the total population (Fig. 1.A). However, there was a difference between groups (70.7 ± 3.3 vs 58.2 ± 10.9; 72.3 ± 5.0 vs 58.2 ± 10.2; 69.4 ± 6.2 vs 57.5 ± 9.2 ml/min/kg, respectively, for the performance group vs the recreational group at a 1, 10, and 25 % slope, $p < 0.05$, Fig 1.A). Similarly, VO2 at ventilatory thresholds were higher in the performance group vs the recreational group both at first ventilatory threshold (VT1) (1% slope: 52.5 ± 3.7 vs 42.3 ± 8.4; 10% slope: 55.0 ± 7.3 vs 41.5 ± 8.5; 25% slope: 51.0 ± 8.7 vs 39.8 ± 8.9 ml/min/kg, respectively, $p<0.05$) and VT2 (1% slope: 65.2 ± 5.6 vs 53.6 ± 10.8; 10% slope: 68.3 ± 7.1 vs 53.0 ± 10.1; 25% slope: 64.0 ± 8.6 vs 50.6 ± 8.9 ml/min/kg, respectively, for the performance and the recreational group, $p<0.05$). No differences at VT1 and VT2 are observed between conditions and groups when oxygen consumption and heart rate values are expressed as relative values (Figure 2).

Figure 1

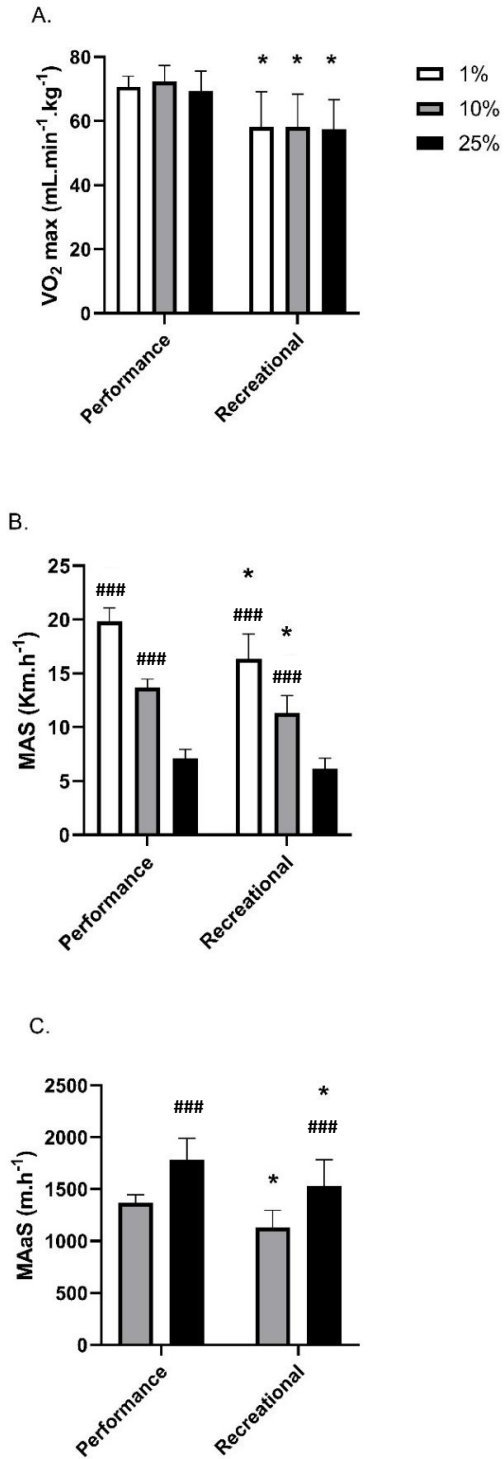

**Figure 1:** Comparison of Maximal oxygen consumption (VO2max), Maximal aerobic speed (MAS) and Maximal Aerobic Ascensional Speed (MAaS) between groups (Performance vs Recreational) and conditions (1-10-25% slope). *, difference between Performance and Recreational p<0.05. ###, difference between conditions, p<0.001.

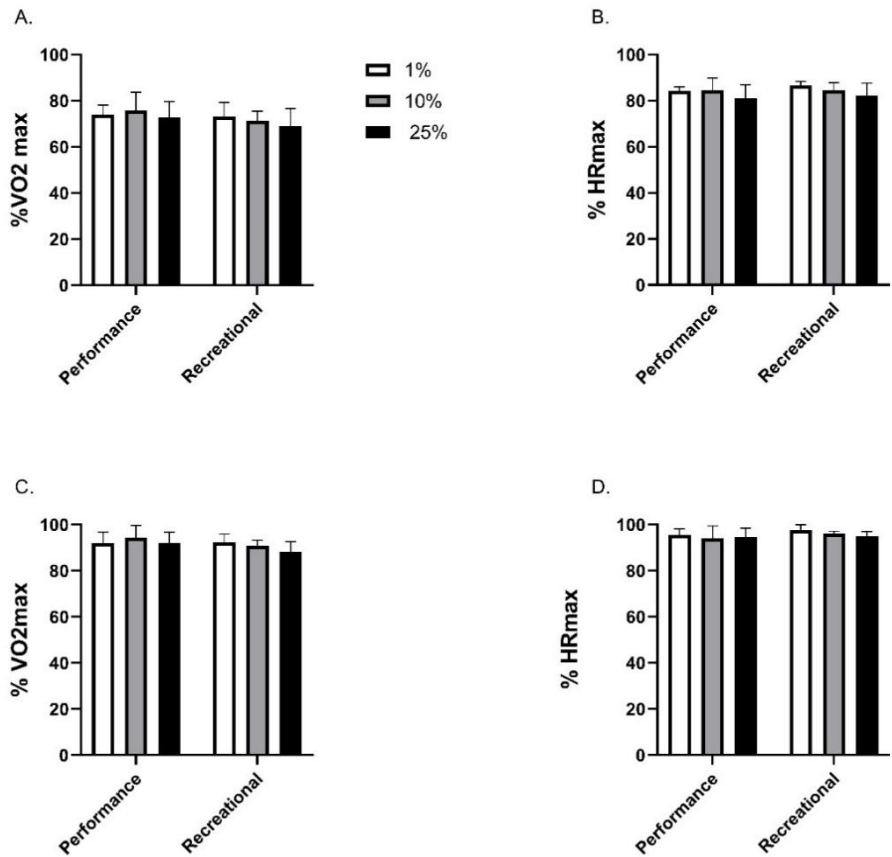

**Figure 2:** Comparison of Maximal oxygen consumption (%VO2max) and Maximum Heart Rate (%HRmax) at Ventilatory threshold 1 (A and B) and Ventilatory threshold 2 (C and D) between groups (Performance vs Recreational) and conditions (1-10-25% slope).

Ascensional speed at ventilatory thresholds are different between conditions ($p<0.001$) and groups ($p<0.05$). The highest ascensional speed was reached at slope of 25%. Interestingly, at ventilatory thresholds, the metabolic parameters, expressed in absolute or relative values, are similar between conditions (10 and 25%) while the ascensional speeds were different (at VT1: 985 ± 107 vs 800 ± 141 and 1300 ± 153 m/h vs 1066,7 ± 285; at VT2: 1257 ± 98 vs 1050 ± 164 and 1614 ± 177 vs 1367 ± 207 for respectively performance vs recreational group at slope 10 and 25%).

No difference between groups or between conditions were observed in the performance of the time to exhaustion test. The mean values for all subjects are 4:36±1:35 (min:sec) at 1%, 3:33 ±1:08 at 10%, and 4:25 ±1:46 at 25%.

Correlation analysis revealed a positive influence of VO2 and ITRA index ($p<0.001$) on MAS and MAaS. The average power recorded over the last three repetitions of our endurance test at 240°/s on the dominant leg has a positive influence on MAS on flat and MAaS at a 10% slope ($p<0.05$) and a 25% slope ($p<0.01$). Fat mass has a negative influence on MAS and MAaS at a 10% slope ($p<0.05$) and at a 25% slope ($p<0.001$). Mixed models showed that the ITRA index, average power at 240°/s on the dominant leg, and VO2max are the best predictor of the MAS at a slope of 1% and the MAaS at the 10 and 25% slope (Table 1).

**Table 1:** Modelling of maximal aerobic speed (MAS) on flat and maximal aerobic ascending speed (MAaS) at a 10 and 25% slope. P.leg: average power recorded over the last three repetitions of our endurance test at 240°/s on the dominant leg. VO2max: maximal oxygen consumption. * : p<0.05.

|  | 1% | 10% | 25% |
|---|---|---|---|
| Intercept | 6.759 * | 353.1 * | 189.1 |
| ITRA index | 0.01311 * | 0.5501 | 0.3302 |
| Leg power | 1.226 | 191.5 | 274.8 |
| VO2max | 0.03989 | 5.753 | 15.34 |
| R² | 0.8873 | 0.9097 | 0.8807 |

## Discussion

The aim of our study was to determine the contribution of cardiorespiratory and muscular parameters to running performance. The major findings were: (i) the slope influences the MAaS, (ii) metabolic parameters at ventilatory thresholds expressed in absolute or relative values are similar between conditions (10 and 25%) while the MAaS are different, and (iii) the predictors of the MAaS were different according to the slope.

The results of the influence of the slope on VO2max are contradictory: some studies observed an influence of slope **(Taylor et al. 1955; Astrand and Saltin 1961; Scheer et al. 2018)** while others did not **(Davies et al. 1974; Kasch et al. 1976; Balducci et al. 2016)**. Our results are in line with the later studies; we do not find any difference in VO2max between the slope conditions. The level of the population and/or the type of test used to determine VO2max may explain the differences across the studies **(Astorino et al. 2004)**.

For example, the results of Scheer et al.'s study are not in line with our study because these authors found differences in VO2max between conditions **(Scheer et al. 2018)**. This study included a homogeneous group of 13 trailers at a good level (VO2max mean over the three tests at 59.8 ± 5.6 mL/min/kg) in contrast to our sample which is rather heterogeneous (VO2max mean over the three tests at 64.9 ± 9.74 mL/min/kg) **(Scheer et al. 2018)**. Moreover, the methodology used to determine the VO2max is different between studies. Indeed, at a 1% slope and during the Trail Test, the increments are different (+2 km/h every 3 minutes in Scheer et al.'s study vs + 1 km/h every minute for our study). The Trail Test of Scheer et al. starts at a 0.5% slope at 10 km/h with an increment of 0.5 km/h and 1% slope gradient every minute, in contrast to our 10% MAS test (fixed gradient, first plateau at 4 km/h and an increment of 1 km/h per minute). Thus, we see that our ascensional speed increment is always fixed, which is not the case for the Trail Test where the increment depends on the number of stops (107m/h to 217m/h vs 100m/h fixed in our study). Above 150 m/h the increment seems too high. However, too large an increment may lead to an underestimation of VO2max 16. In a more recent study by the same author, where the relationship between performance and the same tests as in the 2018 study were investigated, it was shown that the Trail Test was the least predictive of performance on a 31 km trail running course **(Scheer et al. 2019)**. Yet, the Trail Test was the only test where a difference of VO2max had been observed. The accuracy of this test may be questioned. Taken together, therefore, these data suggest that particular attention should be paid to the test protocol used to determine VO2max/MAaS.

The 25% test estimated the highest MAaS, which is in line with the studies showing that a gradient between 20% and a maximum of 30% was necessary to assess MAaS **(Billat 2005; Doucende et al. 2017, 2022)**. Indeed, these same authors point out that above 30%, the restitution of the potential energy of movement by the ankle joint and the Achilles tendon becomes too weak. Below 20%, the horizontal distance becomes too great to reach the highest climbing speeds. Having no difference in VO2max between our tests but different speeds between each condition, it is likely that the energy cost evolves with the slope. This is in line with studies that showed that the horizontal speed is different for the same energy cost for two different slopes **(Balducci et al. 2016; Vernillo et al. 2017)**.

The isokinetic results are in line with the study by Ehrström et al. **(Ehrström et al. 2018)** who observed a correlation between the fatigue index that can reflect the muscular endurance of the knee extensors determined by an isokinetic test of 40 repetitions at 60°/s and performance on a trail-running course. These results do not seem surprising insofar as trail running is associated with a large number of submaximal muscle contractions. Thus, only muscular endurance measurements could be used to objectivise the performance level of the athletes, and should therefore be monitored.

Multiple linear regression models allow us to determine which parameters best explain the MAaS. We selected the following parameters: the ITRA index, the relative average power of the last three repetitions of the 30-repetition endurance test at 240°/s and the condition-specific VO2max (10–25% slope). The weight of the variables varies according to the slope. The muscle power is a parameter that gains in importance with the gradient; a gain of 1 W/kg during the three-repetition endurance test at 240°/sec will lead to an increase of 191 m/h at 10% vs 275 m/h at 25% of the MAaS. The greater influence of muscle strength with the increase of the slope may be explained by the decrease of the potential energy of movement restored by the tendinomuscular elasticity of the ankle joints and Achilles tendon with the increase in slope **(Kasch et al. 1976; Arcelli et al. 2005; Doucende et al. 2017)**. Indeed, the posterior chain is then put under greater stress during an ascent, particularly the ankle, knee, and hip joints **(Doucende et al. 2017; Scheer et al. 2018)**. The multiple linear regression also revealed that for a 1 mL/min/kg increase in VO2max, MAaS increases by 5.75 m/h at 10% vs 15.34 m/h at 25%. Therefore, this suggests that the share of VO2max in performance is related to the increase in slope and that VO2max impact MAaS. This could be explained with the increased energy cost of running uphill and a greater contribution from the posterior chain thus increasing the oxygen demand of muscles **(Prilutsky and Gregor 2001)**.

## Conclusion and pratical applications

The emergence and development of multi-sport GPS watches have made it possible to monitor performance over even the longest trail distances. A wide range of instantaneous measurements are now available to runners, including instantaneous ascensional speed. This measurement is based on GPS, accelerometer, and barometric signals of varying degrees of accuracy. Our results show that for the same metabolic load, the MAaS is different, suggesting that physiological variables such as heart rate should be used to prescribe exercise zones rather than MAaS. Given the specificities of the discipline, heart rate appears to be a more relevant measure to evaluate the effort in trail running.

The ascensional speed remains a relevant parameter for training to individualise uphill training sessions, as long as the inclination of the slope is known and fixed. It is possible to rely on the ascensional speed in addition to the heart rate to determine the intensity of the training. For more focussed events with a single climb, such as vertical kilometres (covering 1000m of D+ over the shortest distance, often 3 to 4km horizontal), this value of ascensional speed is a real added value, provided that the measurement of altitude and the recording of horizontal speed are reliable.

## Conflict of interest disclosure

The authors declare that they comply with the PCI rule of having no financial conflicts of interest in relation to the content of the article.

## Funding

This research did not receive any specific grant from funding agencies in the public, commercial, or not-for-profit sectors

## References

Arcelli E, Torre ALA, Alberti G, et al (2005) Biomeccanica della corsa in salita e rischio di infortuni. 2005

Astorino TA, Rietschel JC, Tam PA, et al (2004) Reinvestigation of optimal duration of VO2max testing. Journal of Exercise Physiology Online

Astrand PO, Saltin B (1961) Maximal oxygen uptake and heart rate in various types of muscular activity. Journal of Applied Physiology 16:977–981. https://doi.org/10.1152/jappl.1961.16.6.977

Balducci P, Clémençon M, Morel B, et al (2016) Comparison of level and graded treadmill tests to evaluate endurance mountain runners. Journal of Sports Science and Medicine 15:239–246

Billat V (2005) L'entraînement en pleine nature. Bruxelles

Davies CTM, Sargeant AJ, Smith B (1974) The physiological responses to running downhill. European Journal of Applied Physiology and Occupational Physiology. https://doi.org/10.1007/BF00423214

Doucende G, Chamoux M, Defer T, et al (2022) Specific Incremental Test for Aerobic Fitness in Trail Running: IncremenTrail. Sports (Basel) 10:174. https://doi.org/10.3390/sports10110174

Doucende G, Rissetto C, Mourot L, Cassirame J (2017) Biomechanical adaptation of preferred transition speed during an incremental test in a gradient slope. Computer Methods in Biomechanics and Biomedical Engineering 20:69–70. https://doi.org/10.1080/10255842.2017.1382865

Ehrström S, Tartaruga MP, Easthope CS, et al (2018) Short Trail Running Race: Beyond the Classic Model for Endurance Running Performance. Medicine and Science in Sports and Exercise 50:580–588. https://doi.org/10.1249/MSS.0000000000001467

Kasch FW, Wallace JP, Huhn RR (1976) VO2(max) during horizontal and inclined treadmill running. Journal of Applied Physiology. https://doi.org/10.1152/jappl.1976.40.6.982

Kerhervé HA, Millet GY, Solomon C (2015) The dynamics of speed selection and psycho-physiological load during a mountain ultramarathon. PLoS ONE. https://doi.org/10.1371/journal.pone.0145482

McLaughlin JE, Howley ET, Bassett DR, et al (2010) Test of the classic model for predicting endurance running performance. Medicine and Science in Sports and Exercise. https://doi.org/10.1249/MSS.0b013e3181c0669d

Paavolainen L, Nummela A, Rusko H (2000) Muscle power factors and VO2max as determinants of horizontal and uphill running performance. Scandinavian Journal of Medicine and Science in Sports. https://doi.org/10.1034/j.1600-0838.2000.010005286.x

Prilutsky BI, Gregor RJ (2001) Swing- and support-related muscle actions differentially trigger human walk-run and run-walk transitions. Journal of Experimental Biology 204:2277–2287

Scheer V, Janssen TI, Vieluf S, Heitkamp HC (2019) Predicting trail-running performance with laboratory exercise tests and field-based results. International Journal of Sports Physiology and Performance. https://doi.org/10.1123/ijspp.2018-0390

Scheer V, Ramme K, Reinsberger C, Heitkamp HC (2018) VO 2 max Testing in Trail Runners: Is There a Specific Exercise Test Protocol? International Journal of Sports Medicine 39:456–461. https://doi.org/10.1055/a-0577-4851

Taylor K, Buskirk E, Henschel A (1955) Maximal oxygen intake as an objective measure of cardio-respiratory performance. Journal of applied physiology. https://doi.org/10.1152/jappl.1955.8.1.73

126 Vernillo G, Giandolini M, Edwards WB, et al (2017) Biomechanics and Physiology of Uphill and Downhill
127     Running. Sports Medicine. https://doi.org/10.1007/s40279-016-0605-y

128